\documentclass[journal,comsoc,a4paper]{IEEEtran}
\usepackage{amsfonts}
\usepackage{amsmath}
\usepackage{cite}
\usepackage{array}
\usepackage{graphicx,cite,epsfig}
\usepackage{subfigure}
\usepackage{stfloats}
\usepackage{setspace}
\usepackage{url}
\usepackage{amssymb}
\usepackage{float}
\usepackage{indentfirst}
\usepackage{makeidx}
\usepackage{tabularx,ragged2e}
\usepackage{multirow}
\usepackage{cases}
\usepackage{algorithm}
\usepackage{algorithmic}
\usepackage{mathtools}
\usepackage{bm}
\usepackage{setspace}
\usepackage{makecell}
\usepackage{xcolor}
\usepackage{ntheorem}
\usepackage{booktabs}
\usepackage{bm}

\graphicspath{{./figures/}}

\begin{document}
\title{C-GRBFnet: A Physics-Inspired Generative Deep Neural Network for Channel Representation and Prediction}
\author{Zhuoran~Xiao, 
        Zhaoyang~Zhang, 
        Chongwen~Huang, 
        Xiaoming~Chen, 
        Caijun~Zhong, 
        and~M\'{e}rouane~Debbah}

\maketitle

\begin{abstract}

In this paper, we aim to efficiently and accurately predict the static channel impulse response (CIR) with only the user's position information and a set of channel instances obtained within a certain wireless communication environment. Such a problem is by no means trivial since it needs to reconstruct the high-dimensional information (here the CIR everywhere) from the extremely low-dimensional data (here the location coordinates), which often results in overfitting and large prediction error. To this end, we resort to a novel physics-inspired generative approach. Specifically, we first use a forward deep neural network to infer the positions of all possible images of the source reflected by the surrounding scatterers within that environment, and then use the well-known Gaussian Radial Basis Function network (GRBF) to approximate the amplitudes of all possible propagation paths. We further incorporate the most recently developed sinusoidal representation network (SIREN) into the proposed network to implicitly represent the highly dynamic phases of all possible paths, which usually cannot be well predicted by the conventional neural networks with non-periodic activators. The resultant framework of Cosine-Gaussian Radial Basis Function network (C-GRBFnet) is also extended to the MIMO channel case. Key performance measures including prediction accuracy, convergence speed, network scale and robustness to channel estimation error are comprehensively evaluated and compared with existing popular networks, which show that our proposed network is much more efficient in representing, learning and predicting wireless channels in a given communication environment. 

\end{abstract}


\section{Introduction}
\subsection{Motivation}
\IEEEPARstart{I}t is crucial for the different parties of a communication system to have accurate channel state information (CSI) so as to achieve optimal system design and deployment. However, efficient and accurate CSI acquisition is very challenging in a wireless context due to the complicated scattering environment which usually contains numerous unknowns and thus leads to unaffordable signaling overhead. 

The common way to acquire CSI is by using pilot symbols or a training sequence. In general, CSI can be obtained through two classes of approaches. One is statistic model-based channel estimation, in which the model is often statistically defined by a known function with a certain number of tuning parameters. When the parameters used to represent the channel are not enough, the accuracy of CSI estimation is hard to guarantee. In contrary, when the channel parameters are in large number, a lot of pilots are needed to estimate them that may lead to high signaling overhead and processing delay. The other type of approaches which can be called as deterministic channel synthesis, like the ray-tracing approach, is based on the specific electromagnetic (EM) wave propagation within a real environment w/wo additional modification from channel measurements. Owing to the analytical framework rooted from electromagnetic theory, this method considers the field properties of the antennas and the propagation paths and can achieve favorable accuracy with excellent cost-and-time efficiency, despite of its full knowledge requirement of the real environment which is usually hard to know in practice.

Note that in a specific communication scenario, a base station always serves a certain fixed area in which the number and locations of the scatterers are also fixed. The wireless channel is then largely determined by the scatterers distributed therein and the corresponding multi-path propagation behavior of EM waves such as specular reflection, diffraction, diffusion, blocking, etc. Although there are inevitable variation in channel characteristics due to the small-scale perturbation of transmission media, the long-term multi-path components of a channel can be regarded quasi-static given the scatterers within that area do not obviously change. Moreover, all the wireless communication processes taking place in a given area are influenced by the same scattering environment. Thus, the static channel characteristics in different locations are highly correlated. So, it is reasonable that we can predict the static CSI of a user equipment (UE) located at a certain point in a fixed area with a set of sample CSIs measured randomly within that area. As a matter of fact, the communication processes take place all the time in the served area for a base station, producing huge amount of channel measurements with corresponding locations that can be potentially made use of and thus leading to significant reduction of signaling overhead and measurement cost. Therefore, it is reasonable to explore such static channel characteristics and combine it with the dynamic channel characteristics instantaneously estimated with pilots to obtain a complete and accurate channel state information. This inspires us to study the prediction of static channel characteristics within a specific communication environment. 

As evidenced by many applications in recent years, deep neural network (DNN) has a strong capability of establishing functional relationship between the input and the output data automatically. We just resort to deep neural network to solve the question we posed above. Basically, our goal is to design a learning machine which can automatically learn the fixed scattering environment and propagation characteristics of EM wave. The learning machine only takes the user's location coordinates as the input, and all other information such as the position of the base station, the scatterers' locations and EM characteristics in the working frequency band are ultimately learned as hidden knowledge represented by the parameters of the neural networks. Sampled channel impulse response (CIR) is chosen as the output of learning machine for intuitive description of the static multi-path characteristics of the wireless channel. 

However, the above task is by no means trivial since it needs to reconstruct the high-dimensional information (here the CIR everywhere) from the extremely low-dimensional data (here the location coordinates), in contrary to its reverse problem of location prediction based on the channel state information. Conventional learning networks either require a huge amount of training data, or result in overfitting and large prediction error. Moreover, the phases of all multi-path EM waves are highly sensitive to the spatial displacement, which has an ultimate impact on the resultant CIR. Conventional DNNs mainly aim to track the change of CIR with respect to the offset of the location coordinates of the user itself (in a way conventional function fitting usually does), which can neither well reflect the overall characteristics of all the different propagation paths nor adequately track their highly dynamic phase changes. How to explore and exploit the information hidden in the physical processes of EM wave propagation which results in the ultimate channel characteristics to design an efficient channel representation and prediction network is thus our main objective in this paper.

\subsection{Related Work}

Comprehensive works has done in CSI prediction in the past few years. In \cite{554747}, the RBF-NN was used to predict the path loss. The transmitter (TX) and receiver (RX) heights, TX-RX distance, carrier frequency, and intercede range were the inputs, and the output was the path loss. In \cite{6748900}, the FNN and RBF-NN were combined with ray launching in complex indoor environments. The neural network was used to predict the intermediate points in the ray launching algorithm to decrease the computation complexity. In \cite{8932272,8116491}, the neural networks were used to remove the noise from the measured CIR, and the PCA was utilized to exploit the features and structures of the channel and model the CIR. In \cite{7590098} the MLP was applied to predict the received signal strength. TX-RX distance and diffraction loss were the inputs, while the output was the received signal strength. In \cite{2016OnThe}, the SVM was used to predict the path loss for smart metering applications. It was based on received signal strength measurements and a 3-D map of the propagation environment.

Most existing works on channel prediction incline to treat it as a sequence prediction problem. Through statistically modeling a wireless channel as a set of radio propagation parameters, two conventional prediction approaches, namely parametric model \cite{6945858} and auto-regressive model \cite{841729} have been proposed. Due to the gap between the model and real wireless channel, the potential achievable prediction accuracy is generally unfavorable which limits their real application. Some recent works have made some attempts in applying machine learning algorithms by making use of its capability on time-series prediction. In \cite{2002Recurrent,6755477}, recurrent neural network (RNN) is proposed to build a predictor for narrow-band single-antenna channels. And RNN is further replaced by a long short-term memory (LSTM) and gated recurrent unit (GRU) in \cite{9128426}. In \cite{8395053,9128426,9277535}, sequential channel state information is viewed as image and CNN+LSTM structure is used for channel predicting. In \cite{8979256}, the learning structure CNN+AR is compared with LSTM. In \cite{8904286}, seq2seq network is applied. However, there are at least two main drawbacks of those methods. From the perspective of data obtaining, a sequence of measured channel with equal sampling time or distance is needed causing high signaling overhead. From the perspective of learning structure, predicted channel is output by a deep neural network in a universal, purely data-driven way without making full use of the implicit physics feature of CSI data. Thus, a huge amount of training data is needed for those learning machine, which is not practical in real scenarios.

There are some related works where radio propagation environment information of typical propagation scenarios is used to apply prior information to help model a wireless channel and reduce the cost used to determine the CSI \cite{8932164,8968729}. This kind of channel model is called tomographic channel model. The main difference between this model and traditional stochastic model is that some of the model parameters are no longer obtained through channel estimation but considered static. Those parameters include the number of multipath, the angle of arrival and the time delay. Existing works have proposed some methods to obtain those parameters, such as the clustering methods \cite{8932164} and machine learning based methods \cite{8968729}. However, all the existing related works can only get part of the channel information. For example, the machine learning based method use the position information as the input of a deep neural network but cannot output the full CSI but some parameters such as the loss factor, angle of arrival and number of paths. Pilots are still needed for obtaining full CSI data.

\subsection{Contributions}
In this paper, we resort to a generative framework and bring out the so-called Cosine-Gaussian Radial Basis Function network (C-GRBFnet) to represent and predict the static channel characteristics in a given environment.  By mathematical analysis of the physical process of CIR formulation, it can be found that the amplitude of each propagation path is approximately determined by the distance between the destination and the image of the source reflected by the corresponding scatterer, while its phase is nearly periodic along the path. Therefore, in the proposed framework, a DNN is used to find all possible images of the sources induced by the scatterers, and an RBF network is used to approximate the amplitudes of the corresponding propagation paths from the images to the designation. Furthermore, inspired by the most recent developed sinusoidal representation network (SIREN) which is able to achieve high-resolution reconstruction for physical signals with fine details on a pixel-by-pixel basis with only the coordinates as input - a problem similar to ours that needs to reconstruct channel with only location coordinates, we propose to incorporate SIREN into RBF network by adding a periodic part to the original Gaussian kernel function in order to track the phase variation across the space. 

For simplicity and convenience, we first illustrate our model setting and mechanism of the proposed methods for a SISO regime. Considering that scaling the number of antennas up is crucial to current and future wireless system \cite{8861014,7400949,6736746,6824752}, we also propose two approaches to extend our methods to a multi-antenna regime through making use of the spatial correlation of antennas.

To summarize, our main contributions are as follows:
\begin{itemize}
	\item 
	We design a novel physics-inspired generative framework to implicitly represent and predict the static channel characteristics, which uses a DNN to find all possible images of the sources induced by the scatterers and an RBF network to approximate the amplitudes of the corresponding propagation paths from the images to the designation, and incorporates SIREN into the framework to track the phase variation across the space.
	
	\item The resultant Cosine-Gaussian Radial Basis Function network (C-GRBFnet) can be properly trained with all past CSI instances obtained in the interested environment, and provide efficient and accurate prediction of the static CIR at each location within that environment with less training data and better convergence performance compared to the conventional learning architectures.
	
	\item The proposed framework is also extended to the multi-antenna regime. Key performance measures, including prediction accuracy, convergence speed, network scale as well as robustness to channel estimation error, are comprehensively evaluated and compared with existing popular networks, which show that our proposed network is much more efficient in representing, learning and predicting wireless channels in a given communication environment. 
\end{itemize}

The remainder of this paper is organized as follows. The physics process of CIR formulation is introduced and why conventional DNN does not work well for the considered problem is analyzed in section \ref{chap:Model}. The design of the physics-inspired network and how to extend it to the MIMO system is given in section \ref{chap:structure}. Section \ref{chap:datasets} introduces our experiment setup and the design of the training datasets. Numerical evaluation from different perspectives are provided in section \ref{chap:results}. Section \ref{chap:mimo} reports the experiment results of extending our methods to MIMO system by comparing to existing works. Section \ref{chap:conclusion} draws the conclusion.

\section{The CIR Formation and Representation}
\label{chap:Model}
In this section, we first introduce how CIR is generated from the perspective of physical EM-wave propagation process. From the perspective of physics and the related mathematical expressions, the prediction problem we proposed can be viewed as an implicit function approximation problem which is suitable for adopting machine learning methods to solve it. However, as we discussed in the sequel, traditional learning structures cannot work well in solving this sort of problem.

\subsection{CIR Formation from a Perspective of EM-wave Propagation} 
\label{chap:CIRFormation}
Physically, the representation of CIR can be accomplished by an analytic approach which decomposes it into a sufficiently large number of separate line-of-sight (LOS) and non-line-of-sight (NLOS) propagation paths, similar to the Ray-Tracing modelling \cite{6815890}. For simplicity, only the LOS path and the specular NLOS paths are considered in our analytical model because these two kinds of propagation paths usually dominate the resultant CIR and provide the vast structural information for a learning neural network. Therefore, given the scatterer distribution in a specific environment, the CIR between an arbitrary TX and a fixed RX can be conventionally denoted as
\begin{equation}\label{CIRfunc}
	h(f, \tau ) = \sum\limits_{k = 1}^N {{a_k}\delta (\tau  - {{{d_k}} \over v_c}){e^{ - j2\pi f{{{d_k}} \over v_c}}}},
\end{equation}
where $N$ is the maximum number of propagation paths, ${{a_k}}$ is the loss factor of $k$th path, $f$ is the carrier frequency, $v_c$ is the speed of light, and $d_k$ is the path length. In the case that the paths are not separable and are continuously distributed which are hard to be represented, we need to sample the CIR function with adequately small time and frequency intervals so that at most only one arrival path can be found in every sampling segment. As we aim to predict the CIR for a real channel without explicitly recovering the CIR function parameters but instead, by implicitly representing it with a universal neural network with only the location of TX as input, we first explore the intrinsic structural feature of the CIR in the sequel.   

\begin{figure}
	\centering
	\includegraphics[width=0.32\textwidth]{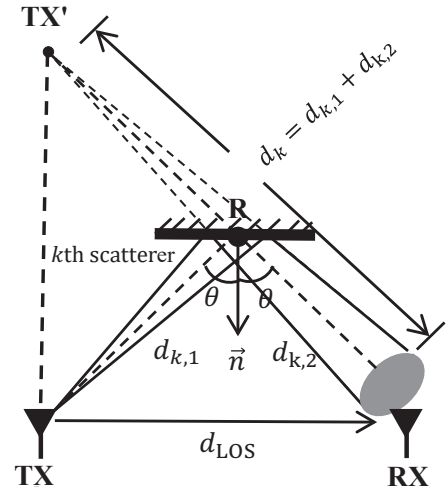}
	\caption{EM-wave propagation along the LOS path and the specular reflection along the $k$th path.}
	\label{RF}
\end{figure}

As shown in Fig. \ref{RF}, according to the Friis Equation, the received signal power of the LOS path can be written as
\begin{equation}
	{P_\textrm{RX}}({d_\textrm{LOS}}) = {P_\textrm{TX}}{G_\textrm{TX}}{G_\textrm{RX}}{({\lambda  \over {4\pi {d_\textrm{LOS}}}})^2},
\end{equation}
where ${d_\textrm{LOS}}$ denotes the length of the LOS path, ${P_\textrm{TX}}$ the transmit power of TX, ${G_\textrm{TX}}$ the transmit antenna gain, ${G_\textrm{RX}}$ the receive antenna gain and $\lambda=c/f$ is the wavelength of the EM wave. The EM field strength at RX of LOS path can be written as
\begin{equation}
	E = {{\sqrt {{P_\textrm{TX}}{G_\textrm{TX}}} } \over {{d_\textrm{LOS}}}}({\lambda  \over {4\pi }}).
\end{equation}
Thus, the LOS path loss factor ${a_\textrm{LOS}}$, i.e., the square-root of the ratio of the received power to the transmit power when all the antenna gains are set to $1$, is
\begin{equation}\label{lossLOS}
	{a_\textrm{LOS}} = {\lambda  \over {4\pi {d_\textrm{LOS}}}}.
\end{equation}

As for the specular NLOS paths, we can see also from Fig. \ref{RF} that the EM waves received at RX are mainly determined by the specular images of the TX antenna reflected by all the scatterers within the environment. For illustration, the specular image TX' of TX produced by the $k$th scatterer can be viewed as a new virtual EM source that radiates the EM wave from TX' to RX instead of TX itself, as shown in Fig. \ref{RF}. If the EM wave experiences multiple reflections among the scatterers (i.e., high-order reflection) along the path before arriving the RX antenna, such images still exist and can be determined by simply finding out the image of images, and so on. In this case, the resultant path length can be written as ${d_k} = \sum\limits_i {{d_{k,i}}} $ and the propagation delay can be written as ${\tau _k} = {d_k}/c$, with $d_{k,i}$ being the length of the $i$th segment of path $k$.

For the simple first-order reflection scenario, the EM field strength received at RX can be expressed as
\begin{equation}
\left\{
\begin{array}{lr}
{E_{\textrm{RX},\bot} = E_{\textrm{TX},\bot} \cdot {R_\bot } \cdot \frac{1}{{d_{k,1}} + {d_{k,2}}} \cdot {e^{ - j2\pi {{{d_{k,1}} + {d_{k,2}}} \over \lambda }}}} \\
{E_{\textrm{RX},\parallel} = E_{\textrm{TX},\parallel} \cdot {R_\parallel } \cdot \frac{1}{{d_{k,1}} + {d_{k,2}}} \cdot {e^{ - j2\pi {{{d_{k,1}} + {d_{k,2}}} \over \lambda }}}}
\end{array},
\right.
\end{equation}
where
$E_{\textrm{RX},\bot}$ and $E_{\textrm{RX},\parallel}$ represent the vertical and horizontal polarization component of the reflection wave field strength at RX, respectively.
And ${R_\bot}$ and ${R_\parallel}$, the reflection factor at the intersection point $\textrm{R}$ on the scatterer, can be written as
\begin{equation}
	\left\{
	\begin{array}{lr}
		{R_ \bot } = {{\cos \theta  - \sqrt {\varepsilon  - {{\sin }^2}\theta } } \over {\cos \theta  + \sqrt {\varepsilon  - {{\sin }^2}\theta } }} \\
		{R_\parallel } = {{\varepsilon \cos \theta  - \sqrt {\varepsilon  - {{\sin }^2}\theta } } \over {\varepsilon \cos \theta  + \sqrt {\varepsilon  - {{\sin }^2}\theta } }}
	\end{array},
	\right.
\end{equation}
in which $\theta $ is the incident and the reflection angle, $\varepsilon $ is the equivalent electrical parameters of the reflection surface. For simplicity, assuming that the electromagnetic wave is purely horizontal polarized. Therefore, the loss factor for the $k$th specular NLOS path can be written as
\begin{equation}\label{lossNLOS}
	a_{\textrm{NLOS},k} = 
	{\lambda  \over {4\pi d_k}}({{\cos \theta  - \sqrt {\varepsilon  - {{\sin }^2}\theta } } \over {\cos \theta  + \sqrt {\varepsilon  - {{\sin }^2}\theta } }}).
\end{equation}

\textbf{\textit{Remark 1:}} The conventional representation of CIR as given by (\ref{CIRfunc}) in fact reveals the structural information of the function itself. It tells us that the CIR at a specific location can be decomposed to more or less EM wave propagation paths, which are further determined by the common scatterers residing in the environment. This not only indicates that it is possible to predict the CIR function by learning the underlying environment structure, but also provides an implicit but more efficient way to represent the channel by exploiting the structure of its component paths.

\textbf{\textit{Remark 2:}} From Equation  (\ref{CIRfunc}), (\ref{lossLOS}) and (\ref{lossNLOS}) we can see that each path (either LOS or NLOS) is featured by a loss factor inversely proportional to the path length and a phase shift proportional to the path length. Note that the path lengths are determined by the relative distances between the virtual transmitters (images of the source or images of the images) and the fixed RX, which makes the CIR function exhibit certain kind of radial and periodic features for the amplitudes and phases respectively. This may call for special considerations in the design of the learning network architecture.       

\begin{figure}[!htb]
	\centering
	\subfigure[The dynamic phase variation of the impulse response of a certain path in a small area ]{\label{phasechange}
		\includegraphics[width=0.4\textwidth]{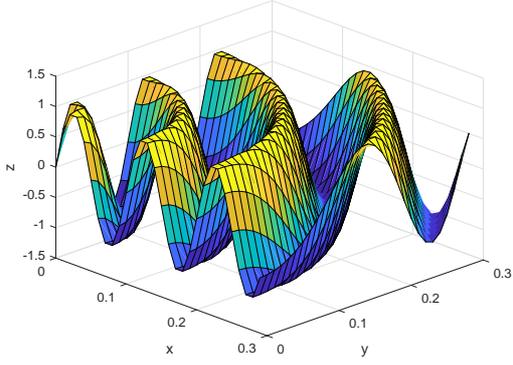}
	}
	\centering
	\subfigure[Approximation result by regular deep neural networks with limited sampling data] {\label{phasesimulation_tanh}
		\includegraphics[width=0.4\textwidth]{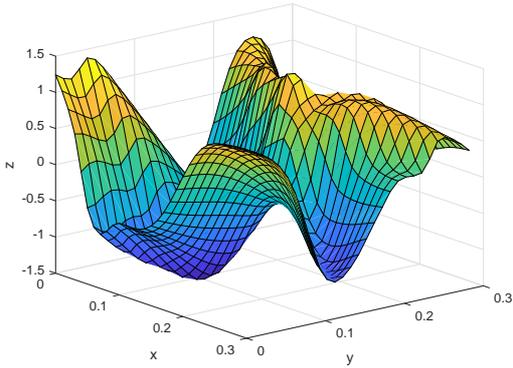}
	}
    \centering
	\subfigure[Approximation result by SIREN neural networks with limited sampling data] {\label{phasesimulation_siren}
		\includegraphics[width=0.4\textwidth]{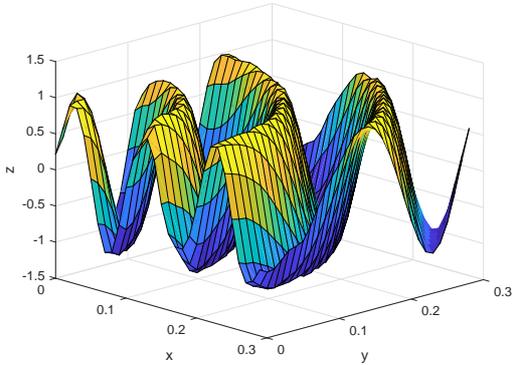}
	}
	\caption{Illustration of the approximation results of the phase part of CIR by different network structures.}
	\label{phase}
\end{figure}

\subsection{Why Conventional DNN Does NOT Work Well}
By analyzing the physics process of EM wave propagation and the mathematical expression of CIR, it can be found that path response is determined by the wave's traveling distance and the reflection angle when reflection exists. Further, the traveling distance and the reflection angle are function of the position of user when the propagation environment is fixed. Thus, we can view CIR as an implicit function of user's position coordinate.

Usually, conventional DNN can work as a universal function approximator and has been proven to perform well in many applications \cite{HORNIK1989359}. However, it fails to do the approximation in this unique problem as observed by our comprehensive experiment attempts. We hereby give a heuristic explain for this phenomenon and show why a specially designed physics-inspired learning structure is needed.

One obvious challenge of our prediction task is that the scattering environment is quite complex and there exist a large scale of unknown implicit parameters to be learned. But a more important issue is related to the sampling density and amount of training dataset. As analyzed in the previous subsection, each component path response of the whole CIR can be decomposed into two parts. One is the amplitude part which is determined by the propagation loss factor and the reflection loss factor. The other one is the periodic phase part caused by the traveling distance accumulation. The propagation path loss is determined by the propagation distance and the reflection angle. As an implicit function of the user's position, the function value changes smoothly with the position coordinates for the reason that the traveling distance of each propagation segment is relatively long enough in real world that a small change of user's position have little impact on the propagation distance or the reflection angel. However, for the phase part, things are different. Because the EM wave length is quite short, the position shift crossing the area will lead to a drastic change of phases and the gap of phase between two adjacent sample points may include several periods of EM wave as shown in Fig. \ref{phasechange}. Thus, the unsmoothness of the phase part brings difficulty for learning-based function approximation if there is no special consideration in the design of network structure.

Smoothness across dataset is significant for the final fitting accuracy for a function approximation problem in machine learning mainly because of the approximation mechanism of using the increasing activation function such as Sigmoid, Tanh and ReLU, etc. The mechanism of these activation functions is basically piece-wise function fitting \cite{1989Approximation}. This mechanism leads to the result like this: for a function approximation task, those prediction values of the points with input coordinates lie between several adjacent training samples will also lie around the values of those sample with smooth change. This phenomenon can be roughly and intuitively shown by Fig. \ref{phasesimulation_tanh}. So, it requires huge number of dataset and a large scale of hidden parameters for a traditional DNN to do such a prediction task. However, it is not practical for a base station to obtain and store a huge dataset and waste a lot of computation resources to finish training such a network. Thus, a specially designed learning structure which has the ability to make full use of physical characteristic of the EM wave propagation is needed. As elaborated below, by properly designing the network structure and its activation function, we can achieve excellent approximation result as shown in Fig. \ref{phasesimulation_siren}.

\section{The Proposed Learning Structure} \label{chap:structure}
In this section, SIREN is introduced and the idea of periodic activation function is illustrated which is suitable for solving our problem. Then, a novel C-GRBF network that combine the GRBF network and the periodic activation function is proposed.
\subsection{SIREN based Generative Network with Periodic Nonlinearity}
\subsubsection{Introduction to SIREN}
the sinusoidal representation networks or SIREN is a special kind of fully connected feed forward neural networks with periodic activation function \cite{SIREN}. The SIREN network can be written as
\begin{equation}
	\Phi (x) = {W_n}({\phi _{n - 1}} \circ {\phi _{n - 2}} \circ  \cdot  \cdot  \cdot  \circ {\phi _0})(x) + {b_n},
\end{equation}
\begin{equation}
	x \mapsto {\phi _i}({x_i}) = \sin ({W_i}{x_i} + {b_i}).
\end{equation}
Here, $ \circ $ represents certain connection between layers. ${\phi _i}:{R^{{M_i}}} \mapsto {R^{{N_i}}}$ is the $i$th layer of the network. It consists of the affine transform defined by the weight matrix ${W_i} \in {R^{{N_i} \times {M_i}}}$ and the biases ${b_i} \in {R^{{N_i}}}$ applied on the input ${x_i} \in {R^{{M_i}}}$, followed by the sine nonlinearity applied to each component of the resulting vector. In fact, periodic nonlinearity have been investigated repeatedly over the past decades, but have so far failed to robustly outperform alternative activation function \cite{7185449,Ren1997Using,1994Improvement,1047806,2019Fourier}. However, a special principled initialization and generalization schemes for the network was proposed in \cite{SIREN} which ensures the convergence of the neural network. 

One important motivation for us to introduce this network to our problem is that it performs well as a coordinate-based neural network for image representation in computer vision \cite{FourierFeatures,NeRF,8953765}. Those works try to reconstruct a picture with pixel coordinate as input and corresponding RGB value of that pixel as output which is similar to our task in some aspects, e.g., both tasks are generative and few samples are available for the network training, and both use low-dimensional input to predict high-dimensional output.

\subsubsection{Approximation Mechanism of SIREN}
Several examples in \cite{SIREN} have demonstrated that SIREN has a better performance in approximating an implicitly defined function with certain characteristic. Here in our problem, the characteristic is named as partial latent periodicity. The formulation of the function can be expressed as
\begin{equation}
	\Phi (x) = F(x)G(\varphi (x)),
\end{equation}
where $F(x)$, $\varphi (x)$ are normal functions of $x$, $G(\varphi )$ is a periodic function of $\varphi$. Thus we call $G(\varphi (x))$ a latent periodic function of $x$ and the $\Phi (x)$ has the characteristic of partial latent periodicity.

According to what we analyzed in last section, neural networks with increasing activation function performs poorly in approximating such function when the frequency of the periodic part is high and the sampling density is relatively low. However, experiments have proven that SIREN network has a strong ability in tracking the latent periodicity of the implicit function through a dataset with relatively low sampling density benefited from its periodic activation function, as shown in  Fig. \ref{phasesimulation_siren}.

Why this kind of neural networks perform so well in certain implicit representation problems is still open. Here we try to give some heuristic explanations with respect to our own problem. There are two features of the networks that make it a universal function approximator. One is that the approximation mechanism of periodic activation function is different from those increasing activation function, like Sigmoid, Tanh and ReLU. The output of every neuron is the sum of a series of sine functions which work very similar to the Fourier neural network \cite{831544} that approximates a function by approximating its Fourier series. The other feature of SIREN is that the activation distributions stay consistent from layer to layer and the gradient statistics also stay consistent from layer to layer \cite{SIREN}. Thus, the network can be designed deep enough with high learning ability without occurring some problems such as gradient explosion and gradient vanishing.

\subsubsection{Network Structure}
As mentioned in the last section, the CIR function is formed by the sum a series of path response. Each path response can be divided into two parts, the fading factor ${a_k}$ and the phase shift ${e^{ - j2\pi{\textstyle{{{d_k}} \over \lambda}}}}$. Given the scattering environment fixed, the fading factor is determined by the position coordinate of user, i.e., 
\begin{equation}
	{a_k} = F({p_x},{p_y},{p_z}),
\end{equation}
and the phase shift part is a periodic function of propagation distance of the path, or 
\begin{equation}
	{e^{ - j2\pi{\textstyle{{{d_k}} \over \lambda}}}} = G({d_k}).
\end{equation}
Obviously, the propagation distance is a function of the user's position coordinates, i.e., 
\begin{equation}
	{d_k} = \varphi ({p_x},{p_y},{p_z}).
\end{equation}
Combined with the above analysis, the CIR function has a latent periodicity characteristic. Thus, the special approximation mechanism of SIREN network fits well with our problem. By utilizing the data implicit periodicity in our learning machine, significantly less data is required. In the meanwhile, by making use of the sine function as the activation function, we can design a network as deep as we need for sufficient learning capability for representing a complicated scattering environment.

As a universal function approximator, it is quite simple to be implemented in practice. We use the user's position coordinate as the input and the vectorized CIR sampling series as the output. The initial scheme proposed in \cite{SIREN} also needs to be adopted, and the function can still be trained by back propagation.

\begin{figure}[htp]
	\centering
	\includegraphics[width=0.5\textwidth]{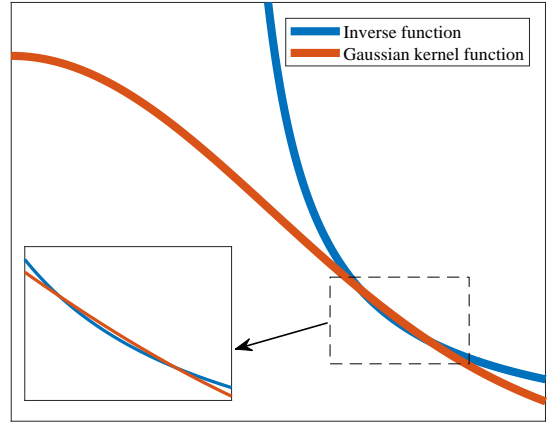}
	\caption{The amplitude change of path response can be well fitted by Gaussian kernel function.}
	\label{gaussian_inverse}
\end{figure}

\begin{figure*}
	\centering
    \includegraphics[width=0.8\textwidth]{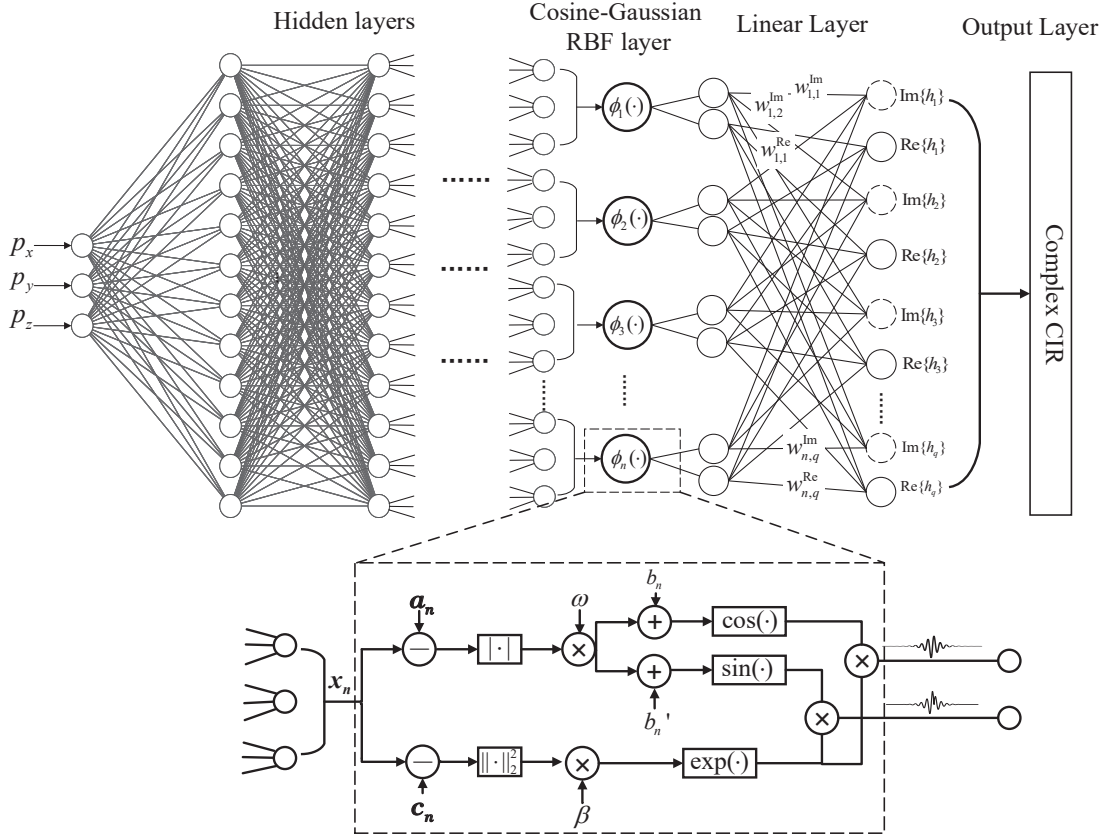}
	\caption{The structure of the proposed C-GRBF network.}
	\label{network}
\end{figure*}

\subsection{C-GRBF: A Physics-Inspired Generative Network}
\subsubsection{Physics-Inspired Neural Network Design}
According to what we analyze in \ref{chap:CIRFormation}, the physics process of CIR formulation can be composed into three stages: path searching, calculation of path response and response synthesis. The response synthesis is naturally done for the reason that our target output of the learning machine is a vector sampled by the time coordinate and at most one arrival path is contained in an interval. The learning structure design is motivated by simulating the physics process of CIR formulating mentioned.

The path searching process is actually a process of finding out all possible images of original user's position coordinates for all involved scatterers at which the reflection happens. Once the images are obtained, the channel parameters such as traveling distance, reflection times and arrival angle are determined. A normal feed forward neural network can work perfectly for this task, by using the historic channel sample dataset as the training input, given the scattering environment fixed.  

The path response is composed by its amplitude part and phase part. As mentioned above, a neural network with periodic activation function can approximate this kind of function well but needs a deep structure. In the meantime, neural network with periodic nonlinearity converges slower than normal networks universally. Thus, we propose using a physics-inspired radial basis function network for this task. 

Radial basis function (RBF) networks have been proved being capable in approximating any function of finite discontinuities with only one hidden layer and thus converge faster than a deep network \cite{2014Universal}. The hidden layer consists of an array of computing units. Each unit containing a parameter vector is called a center, and the unit calculates a squared distance between the center and the network input vector. The squared distance is then divided by a parameter called width and the result is then passed through a nonlinear function. The output layer is essentially a linear combiner with a set of connection weights. Gaussian kernel function is commonly used as the computing unit and can be written as
\begin{equation}
\phi '(\bm{x}) = \exp ({{ - |\bm{x} - {\rm{ }}\bm{c}{|^2}} \over {2{\sigma ^2}}}),
\end{equation}
where ${{\bm{c}}}$ is the center of the kernel and $\sigma$ is a learnable parameter.

There are at least two motivations for us to apply RBF here. First of all, as seen in (\ref{CIRfunc}), (\ref{lossLOS}) and (\ref{lossNLOS})), the phase accumulation is proportional to the propagation distance and the amplitude change is inversely proportional to the total propagation distance , i.e., ${amplitude_{}} \propto {1 \over {d_k}}$. An inverse function is quite suitable for been approximated by a series of Gaussian kernel function especially when the range of $d_k$ lies far from the coordinate origin. As what we show in Fig. \ref{gaussian_inverse}, only a single Gaussian can fit the inverse function well within a certain region. In our work, the region represents the range between the minimum propagation distance to a maximum propagation distance. Actually, the amplitude part is approximated by a set of kernel functions with higher accuracy. Second, as we analyzed, a periodic part must be added to the learning structure for tracking the hidden periodic feature crossing data samples. In our network, we add a cosine function with random initial frequency and a new kernel center as the periodic part. Benefited from the special characteristic of RBF network, it can be directly multiplied with Gaussian function to formulate a new kernel function without losing its differentiability for back propagation training which promises good convergence. The new kernel function is named as Cosine-Gaussian kernel function and is expressed as 
\begin{equation}
	\phi (\bm{x}) = \cos (w|\bm{x} - \bm{a}| + b)\exp (\beta |\bm{x} - \bm{c}|^2),
\end{equation}
where $\bm{a}$ represents the center of the periodic function, $\bm{b}$ the phase shift, $\bm{c}$ is the center of the Gaussian function and $\beta  =  - 1/2{\sigma ^2}$, and they all are learnable parameters. Here, the phase shift term is necessary in the cosine function and the reason will be illustrated in a following part.

\subsubsection{Structure of C-GRBF Network}
The complete learning structure of C-GRBF network is shown in Fig. \ref{network}. Aiming at simulating the physics process of CIR formulating shown in (8). The whole structure is composed of three parts: a deep forward neural network, an RBF layer with Cosine-Gaussian kernel function and a linear layer combining the weighted output of kernel functions. The number of kernel function and length of CIR vector are denoted as $n$ and $q$. ${\phi _n}( \cdot )$ represent the $n$th kernel which outputs the real and imaginary part of a complex number. All the parameters shown in Fig. \ref{network} is learnable.

Here, we describe the information flow in the network in a forward order. The input coordination is first fed into a deep neural network and the output the DNN is composed of a set of neuron groups. Each group has three neurons in order to represent a possible image point of user coordinate for a certain propagation path. It is worth mentioning that for the reason that DNN is jointly trained with the following network structure, the output coordinates are not the real image points' coordinates but with a certain offset. For the same reason, several output coordinates may represent the same image point but with different offsets. Every outputted coordinate is then fed into a Cosine-Gaussian kernel function in the radial basis function layer. ${\bm{c}}$ is a learnable center of Gaussian kernel function, and the amplitude part of a certain path can be obtained by getting the weighted sum of outputs of related kernel functions. After sufficient training, the parameter $w$ will converge to a fixed value which is the frequency of the electromagnetic wave dividing the speed of light. In real wireless communication systems, the central frequency is always known as prior information. So, we suggest to initializing the parameter around the central frequency in order to get a faster convergence speed. Parameter ${\bm{a}}$ is designed as kernel center for periodic function which offsets ${\bm{x}}$ in order to simulate the phase shift caused by the travel distance accumulation. Thus, ${\bm{a}}$ can be viewed as the position coordinate of base station with an offset which can be written as $\bm{a}  = \bm{ {{P_\textrm{BS}}}}  + \bm{{{k_\textrm{offset}}}} $, where $\bm{{{P_\textrm{BS}}}} $ represents the coordinate of base station and $\bm{{{k_\textrm{offset}}}}$ represents a certain offset. Then, we have $|\bm{x_n} - \bm{a_n}| = {d_n} + m_n\lambda $, where $d_n$ is the propagation distance of the corresponding path, $\lambda $ is the length of EM-wave and $m_n$ is a certain integer here. Parameter $b$ is designed for deciding whether this kernel is used for fitting the real or imagine part of element in CIR vector. 

The linear layer after the RBF layer plays two roles. First, for the reason that the whole network is trained jointly and all the parameters are adjustable, the generative process of a path response is calculated by several kernel function pairs. A linear layer is needed to fuse the results together in the form of weighted sum. Second, the linear layer separate different path response apart and arrange the output of our network in the form of CIR vector that we regularized.

There are several advantages of this learning structure. For the reason that the mathematical formulation of path response is simulated by the network structure, prior knowledge of input and output data is added for the learning machine which ensures high prediction accuracy. Moreover, the specially designed RBF layer enables the network to track the periodicity with only one layer, making the network converge faster than other networks with periodic activation function.  

\subsubsection{Cost Functions}
Mean square error (MSE) is used as the cost function of all our neural work. Mean square error can be written as
\begin{equation}
	MSE = {1 \over q}{\sum\limits_{m = 1}^q {({y_m} - \mathop {{y_m}}\limits^ \wedge  )} ^2},
\end{equation}
where $q$ is the dimension of output, $y$ is the training label and ${\mathop y\limits^ \wedge  }$ is the prediction value. It represents the average distance between the target value and the prediction value of all the output dimensions.

Normalized mean square error (NMSE) is used to evaluate the prediction accuracy of the testing dataset, which can be written as
\begin{equation}
	NMSE = E({{\sum\limits_{m = 1}^q {|{y_m} - {{\mathop {{y_m}|}\limits^ \wedge  }^2}} } \over {\sum\limits_{m = 1}^q {|{y_m}{|^2}} }}).
\end{equation}
NMSE is an expectation value calculated across the testing dataset. Compared with MSE, NMSE is more convenient for comparison crossing different datasets. Adam optimizer embedded in TensorFlow library \cite{10.5555/3026877.3026899} is used to train the networks.

\subsection{Complexity Analysis}
Assuming that there are $l$ layers in the front fully connected network and the number of neurons of each hidden layer is $\{ {n_1},{n_2}, \cdot  \cdot  \cdot ,{n_l}\} $, the computation of the front fully connected neural network consists of $3{n_1} + \sum\limits_{k = 1}^{l - 1} {{n_k}{n_{k + 1}}} $ times of multiplication and $\sum\limits_{k = 1}^l {{n_k}} $ times of activation using normal activation function. As mentioned above, the input of the following radial basis function is a set of neuron groups and each group contains three neurons. Thus, the number of the group is ${{{n_l}} \over 3}$. Each kernel function works in the same way as an activation function. Thus, the computation cost of the RBF layer is equivalent to ${{{n_l}} \over 3}$ times of activation. If we further assume the dimension of the final output CIR vector is $m$, the cost of the linear layer is then ${{m{n_l}} \over 3}$ times of multiplication. Therefore, the overall computation cost for our proposed C-GRBF is $3{n_1} + \sum\limits_{k = 1}^{l - 1} {{n_k}{n_{k + 1}}}  + {{m{n_l}} \over 3}$ times of multiplication plus $\sum\limits_{k = 1}^l {{n_k}}  + {{{n_l}} \over 3}$ times of activation.

\subsection{Extension to MIMO Systems}
In this part, we will illustrate how to extend our proposed network to MIMO systems. The reason why channel prediction methods proposed above can be extended to MIMO systems is based on the spatial correlation of antenna array. The correlation is obvious from the perspective of physics. Antennas in an array are placed quite close to each other within the same scattering environment, thus the propagation paths of electromagnetic wave for different antennas are similar. Through making use of the correlation of channels, the network scale for a MIMO system can be greatly reduced. Further, most scattering environment information is contained in the MIMO channel sample data, which ensures high prediction accuracy in our proposed network. 

\subsubsection{Extend By Sharing Parameters of Learning Network}
One intuitive way for using our methods in MIMO systems is by setting an independent learning machine for each TX/RX antenna pair. This approach works but may consume lots of computation and storage resources. In fact, lots of parameters of the independent learning machine can be shared for the reason that similar information is contained in those parameters. For example, path searching for different antennas are highly similar. Thus, a large scale of parameters in the fully connected part of C-GRBFnet can be shared. Also, the amplitude of channel responses of different antennas is also similar, thus some parameters in the RBF layer can also be shared. We propose to use a single larger network to output the whole matrix composed of CIR sampling point of the MIMO system and what we need to do is vectorizing the $N \times M$ matrix to an $NM$ dimension vector as the label of the learning machine. Where $M$ represents the number of antennas and $N$ represents the vector length of sampled CIR for a single antenna.

\begin{figure}
	\centering
	\includegraphics[width=0.5\textwidth]{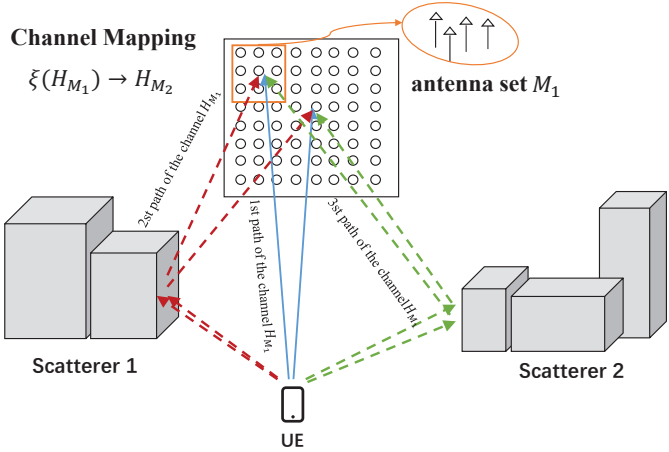}
	\caption{An illustration of the highly correlated EM wave propagation paths from the UE to different antennas. The reason behind this fact is that all the paths are influenced by the same set of scatterers. Thus, there exists a mapping from the channels of a sub-antenna set ${{M_1}}$ to the channels of the whole antenna array ${{M_2}}$.}
	\label{MS}
\end{figure}

\subsubsection{Extend Through Channel Mapping}
By making use of the spatial correlation induced by the similarity of propagation paths from UE to different antennas of an array, the concept of Channel Mapping proposed in \cite{9048929} has been proven to be effective by theoretical analysis and experiment results. As shown in Fig. \ref{MS}, the basic idea of channel mapping is estimating the channels at one set of antennas using the channel knowledge of a different set of antennas. The experiments in \cite{9048929} shows that this mapping even exists in two sets of antennas of different arrays given they reside in the same environment. Thus, this idea can also help to extend our methods to MIMO systems. Firstly, several independent channel prediction learning machines for single antenna in a sub-antenna set ${{M_1}}$ should be trained. It is worth mentioning that the training process can be done in parallel as the number of learning machines is relatively small. In the meanwhile, a deep network which using the CIR matrix composed of the channels of the chosen antenna pairs as input and the CIR matrix of the channel of the whole antenna array ${{M_2}}$ as output can be trained. The architecture of this network can be either a fully connected or a convolutional network. Then, the prediction of the whole channel CIR matrix can be done through inputting the predicted CIR vector of those single antenna pair to the mapping machine. By denoting the mapping as function $\xi ( \cdot )$, the above network works for such a mapping
\begin{equation}
	\xi ({H_{{M_1}}}) \to {H_{{M_2}}},
\end{equation} 
where ${H_{{M_1}}}$ denotes the CIR matrix of the sub antenna set and ${H_{{M_2}}}$ denotes the CIR matrix of the whole antenna array.

\section{Experimental Results and Analysis} \label{chap:results}

\subsection{Scenario Setup} \label{chap:datasets}
In this section, we use the ray-tracing model to simulate an outdoor scenario in real world. Ray-tracing is a deterministic channel model that works through calculating the physics propagation of electromagnetic wave and the real scene is simulated by computer 3D modeling \cite{530804,2019DeepMIMO}.

\begin{figure}
	\centering
	\includegraphics[width=0.5\textwidth]{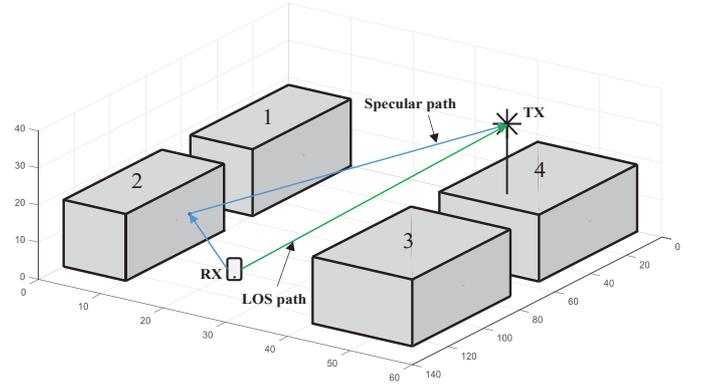}
	\caption{3D model of the target cell}
	\label{scene}
\end{figure}

We choose an outdoor scenario to set up our simulation model. The 3D modeling diagram is shown in Fig. \ref{scene}. In the 3D model, building walls are modeled as rectangular surfaces with specific electromagnetic material properties. This scene is composed of four buildings and a base station. Among them, Building 1 and Building 2 are with the size of $55(x) \times 10(y) \times 18(z)$m, and the two other buildings are with the size of $55(x) \times 16(y) \times 18(z)$m. The distance between Building 1 and Building 2 and the distance between Building 3 and Building 4 are both $16$m. The distance between Building 1 and Building 4 and the distance between Building 2 and Building 3 are both $30$m. The base station locates in ($45$m,$48$m,$37$m). In our simulation, the $y$ coordinate of user's location is set between ($15$m,$30$m) and $x$ coordinate is set between ($20$m,$120$m) in order to guarantee that the propagation paths between the user and base station contain the LOS path and at least one specular path. The height of users is set to $1.6$m. The electromagnetic wave is completely vertically polarized with frequency of $3$GHz.

\subsection{Dataset Generation}
\textcolor{blue}
The progress of generating the datasets is the same with what we state in \ref{chap:CIRFormation}. For the reason that CIR as the target output is a continuous function which is hard to be represented in a learning machine. We need to sample the CIR function at time domain with a sufficiently small interval which is mainly determined by the center frequency and bandwidth of the communication signals and here is chosen to be $1$ns in our setting. 
 Since complex value cannot be directly handled by neural network, we divide the complex number into its real and imaginary components. Then the vectorized representation of CIR function can be written as $\{ \text{real}({y_1}),\text{imag}({y_1}),\text{real}({y_2}),\text{imag}({y_2}), \cdot  \cdot  \cdot ,\text{real}({y_q}),\text{imag}({y_q})\} $ where $q$ is the maximum number of sampling points or the dimension of the CIR vector and it is the target output of our learning machine. 
Here $q$ is determined by not only the sampling interval, but also the sampling duration which is further determined by the maximum delay spread within the area. In our datasets, the sampling takes place within the time slot between $220$ns $\sim$ $310$ns to ensure at least the LOS path and the first-order and second-order specular paths are contained in each CIR instance. Thus, $q$ is $(310-220+1) \times 2 = 182$ (the complex value needs to be split in to real and imagine part). Considering the convergence speed and the robustness of the learning process, we do the data preprocessing and scale the data to a suitable range.

Given a specific area, we use the sampling density to measure the size of datasets. The sampling density is defined as the average number of samples in a unit area ($1m^2$), which represents how many historical data samples do we have stored in the base station. The locations of these samples are randomly generated through a Poison point process. In our experiment setting, three data sampling densities of $40$, $60$ and $100$ are considered. Selection of the three specific sampling density is based on experiment attempts and practically consideration. Practically, the training datasets is obtained from the history records stored in base station. So, we want to achieve a good prediction performance without wasting too much storage capacity of the base station. From the experimental perspective, the performance results under the three sampling densities show significant differences in all the three methods. So, choosing this three sampling densities helps better illustration of important characteristics of the network. It is noteworthy that the sampling density is quite low compared with other works of the field of AI-aided wireless communication based on ray-tracing datasets. For example, 121000 channel samples are used in \cite{9048929} for a $10m \times 10m$ scenery, 100000 channel samples are used in \cite{8395149} for a $40m \times 60m$ area and 50000 samples are used in \cite{8645463} for a $20m \times 30m$ scenery.



\subsection{Model Settings and Training}
\subsubsection{\textbf{Setting of SIREN Network}}
SIREN network can support deeper neural network structure without occurring some problems such as gradient explosion and gradient vanishing. Thus, the tuning of SIREN network is convenient in practical usage.

To find the limitation of the prediction accuracy can be achieved by this method. We use a SIREN network with 5 hidden layers with 380-512-512-512-256 neurons for each layer. We adopt the parameter initial scheme proposed in \cite{SIREN}. It is found that this kind of initial scheme works better than those other most commonly used initial schemes for the convergence speed of SIREN network. The number of training steps is set to 20000. In each step, a mini-batch composed of 20 training samples are used. Training rate is set to $1 \times {10^{ - 3}}$ for better convergence performance. The detailed parameters and values are summarized in Table \ref{SIREN_setting}.

\begin{table}
	\caption{Parameters and Values for SIREN network}
	\begin{center}
		\begin{tabular}{ p{4cm}   p{4cm}}
			\toprule
			\textbf{Parameters} & \textbf{Value} \\
			\toprule
			Input dimension & 3 \\
			
			Output dimension & 182 \\
			
			Activation function & Sin \\
			
			Number of neurons in the hidden layer & 380-512-512-512-256 \\			
			
			Performance metric & Mean square error (MSE) \\
			
			Optimizer & Adam \\
			
			Training steps & $2 \times {10^5}$ \\
			
			Learning rate & $1 \times {10^{ - 3}}$ \\
			
			Batch size & 20 \\
			
			Training samples & $80\% $ of the whole datasets \\
			\toprule
		\end{tabular}		
	\end{center}
	\label{SIREN_setting}
\end{table}

\subsubsection{\textbf{Setting of C-GRBF Network}}
Two experiment attempts will be discussed in this part. To find the limitation of the prediction accuracy can be achieved by this method. We use a deep network composed of 3 hidden layers with 256-512-900 neurons for each layer. For the RBF layer, $300$ kernel function pairs is included. The detailed parameters and values are summarized in Table \ref{RBF_setting}. 

In practice, the tradeoff between prediction accuracy and computation cost is also needed to be taken into consideration. For comparing the performance of C-GRBF network and SIREN network with limited parameters, we constrain the numbers of parameters of both networks to be nearly the same. For the C-GRBF network, the deep network is composed of 3 hidden layers with 128-256-600 neurons for each layer. For the RBF layer, $200$ kernel function is included. The total number of parameters of this network are 192576 including bias items. A SIREN network with 4 hidden layers with 150-256-300-256 neurons for each layer is used for comparison. The total number pf parameters is 193411. For fair comparison, the optimization algorithm, learning rate and training steps are all set to be the same. The dataset is set as sampling density=100.
The detailed parameters and values are summarized in Table \ref{network_comparison}.

\begin{table}
	\caption{Parameters and Values for C-GRBF network}
	\begin{center}
		\begin{tabular}{ p{4cm}   p{4cm}}
			\toprule
			\textbf{Parameters} & \textbf{Value} \\
			\toprule
			Input dimension & 3 \\
			
			Output dimension & 182 \\
			
			Activation function & Sin for the deep part \\
			
			Number of neurons in the hidden layer & 256-512-900 for the deep part and 300 kernel function pairs in RBF layer \\			
			
			Performance metric & Mean square error (MSE) \\
			
			Optimizer & Adam \\
			
			Training steps & $2 \times {10^5}$ \\
			
			Learning rate & $1 \times {10^{ - 3}}$ \\
			
			Batch size & 20 \\
			
			Training samples & $80\% $ of the whole datasets \\
			\toprule
		\end{tabular}		
	\end{center}
	\label{RBF_setting}
\end{table}

\subsubsection{\textbf{Setting of the Benchmark Auto-Encoder Based Learning Structure}}

In order to deal with the problem of high dimension of output, it is intuitive to adopt an Auto-Encoder (AE) for compressing the original CIR vector. Then, a fully connected network is used to learn the mapping of the position coordinates and the low dimension output CIR. Thus, we use such a learning structure as benchmark for comparison with C-GRBFnet. For the Auto-Encoder, a three-layers encoder connected with a three-layers decoder architecture is designed. The number of neurons is almost equal in the two networks. The input vector dimension is 182. A 128-32-3 fully connected networks (FNN) as encoder and a 3-32-128-182 FNN decoder with Sigmoid activation function is designed. After the training of AE and getting the compressed representation of CIR vector, we train a feed forward neural network with the user's position coordinates as input and the compressed CIR vector as label. The detailed parameters and values are summarized in Table \ref{AEtable} and Table \ref{pos-code}.
\begin{table}
	\caption{Parameters and Values for Auto-Encoder network}
	\begin{center}
		\begin{tabular}{ p{4cm}   p{4cm}}
			\toprule
			\textbf{Parameters} & \textbf{Value} \\
			\toprule
			Input dimension & 182 \\
			
			Output dimension & 182 \\
			
			Activation function & sigmoid \\
			
			Number of neurons in the hidden layer & 128-32-3-32-128 \\		
			
			Performance metric & Mean square error (MSE) \\
			
			Optimizer & Adam \\
			
			Training steps & $2 \times {10^5}$ \\
			
			Learning rate & $1 \times {10^{ - 3}}$ \\
			
			Batch size & 20 \\
			
			Training samples & $80\% $ of the whole datasets \\
			\toprule
		\end{tabular}		
	\end{center}
	\label{AEtable}
\end{table}
\begin{table}
	\caption{Parameters and Values for Pos-Compressed CIR network}
	\begin{center}
		\begin{tabular}{ p{4cm}   p{4cm}}
			\toprule
			\textbf{Parameters} & \textbf{Value} \\
			\toprule
			Input dimension & 3 \\
			
			Output dimension & 3 \\
			
			Activation function & tanh \\
			
			Number of neurons in the hidden layer & 128-256-256-128 \\		
			
			Performance metric & Mean square error (MSE) \\
			
			Training steps & $2 \times {10^5}$ \\
			
			Learning rate & $1 \times {10^{ - 2}}$ \\
			
			Optimizer & Adam \\
			
			Batch size & 20 \\
			
			\toprule
		\end{tabular}		
	\end{center}
	\label{pos-code}
\end{table}

\begin{table}
	\caption{Parameters and Values for C-GRBF and SIREN comparison}
	\begin{center}
		\begin{tabular}{ p{4cm}   p{4cm}}
			\toprule
			\textbf{Parameters} & \textbf{Value} \\
			\toprule
			Input dimension & 3 \\
			
			Output dimension & 182 \\
			
			Number of neurons in the hidden layer & C-GRBF: 128-256-600 for the deep part and 200 kernel function in RBF layer. \\
			
			& SIREN: 150-256-300-256 \\			
			
			Performance metric & Mean square error (MSE) \\
			
			Optimizer & Adam \\
			
			Training steps & $2 \times {10^5}$ \\
			
			Learning rate & $1 \times {10^{ - 3}}$ \\
			
			Batch size & 20 \\
			
			\toprule
		\end{tabular}		
	\end{center}
	\label{network_comparison}
\end{table}

\subsection{Experiment Design to Evaluate the Robustness of the Proposed Learning System}
In this part, we aim to design experiments for testing the robustness of our proposed learning structures. Basically, the robustness of neural networks represents the influence on the prediction accuracy when there exist perturbations among training datasets \cite{2009Robustness}. It is also a vital perspective in evaluating a learning system for the reason that the training data obtained through real world measurement is always imperfect.

For our work, the imperfection mainly comes from the measurement error of user's channel impulse response and position coordinates. So, we evaluate the robustness of our proposed learning system through adding some measurements error in our training datasets. For the CIR measurements error, we compare the prediction NMSE change with different measured CIR errors. The CIR measurement error is described by the NMSE of measured channel. Moreover, two kinds of noise distributions are considered. One is Gaussian distribution with zero mean and different deviations. The other is symmetric alpha-stable distribution which is a special case of heavy-tailed distribution and is common in wireless communication systems \cite{6594252,9241843}. In our experiments, the characteristic exponent $\alpha $ is set as $1.2$, $1.5$, $1.8$ separately.

For the position measurements error, the distribution of it is modeled as a Gaussian distribution with zero mean and the standard deviation is set to be 0.03m in our setting for the reason that the positioning error is about 0.02m-0.04m practically. For direct comparison to the results under perfection setting, all the network parameters and values are kept the same and three sampling densities are also discussed.

\subsection{Experiment Design to Evaluate the Generalization Ability of the Proposed Learning System}
In this part, we aim to design experiments for testing the generalization ability of our proposed learning structures. Firstly, in order to evaluate the ability for predicting channel in higher frequency band of our proposed network. We generate several groups of training data under different frequency band while keeping the same parameter settings.

\begin{figure}
	\centering
	\includegraphics[width=0.45\textwidth]{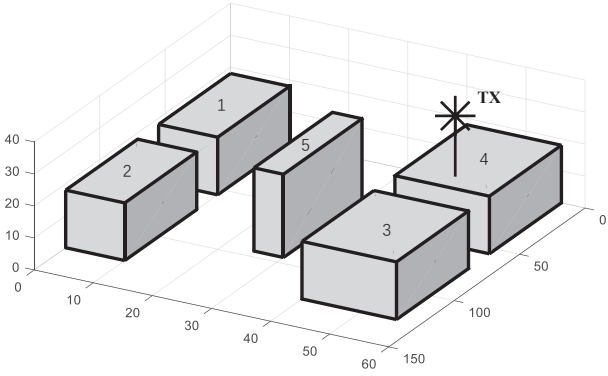}
	\caption{3D model of the scenery for evaluating the performance of the proposed network when there are a lot of NLoS-only samples.}
	\label{environment3}
\end{figure}

Moreover, a new scenery based on the original one is tested to evaluate the performance of the proposed network when there are a lot of NLoS-only samples include. As shown in Fig. \ref{environment3}, in the new 3D model, building 5 with the size of $60(x) \times 5(y) \times 26(z)$m is added in the environment. In this simulation, the $y$ coordinate of user's location is set between ($15$m,$35$m) and $x$ coordinate is set between ($20$m,$120$m) except the area occupied by building 5.  

\begin{figure}
	\centering
	\includegraphics[width=0.425\textwidth]{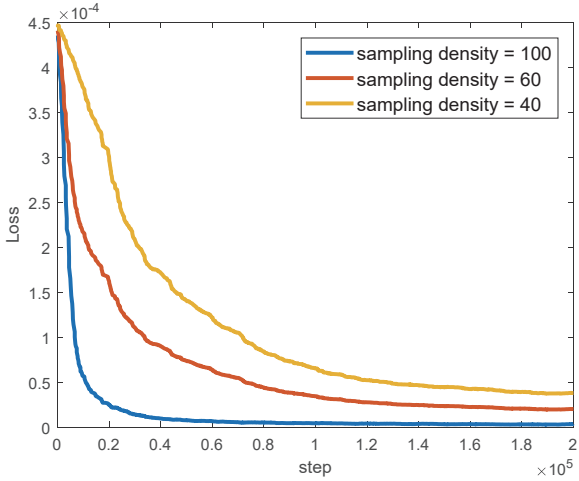}
	\caption{The convergence curve of SIREN network with different sampling density.}
	\label{siren_curve}
\end{figure}

\begin{figure}
	\centering
	\includegraphics[width=0.425\textwidth]{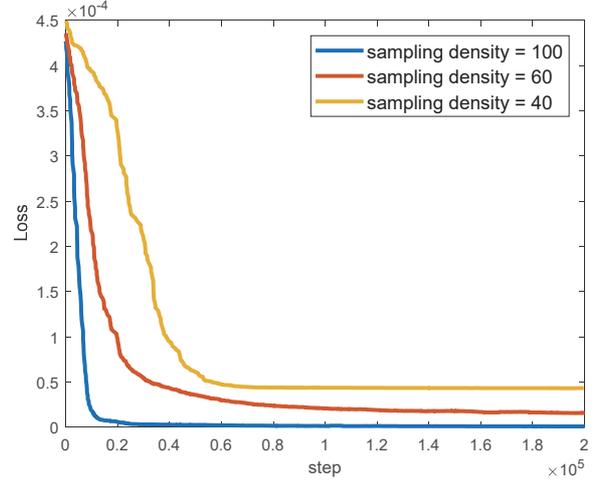}
	\caption{The convergence curve of C-GRBF network with different sampling density.}
	\label{rbf_curve}
\end{figure}

\subsection{Experimental Results and Analysis}
 The convergence curves of SIREN and C-GRBF network are shown in Fig. \ref{siren_curve} and Fig. \ref{rbf_curve}, respectively. Simulation results under three sampling density are illustrated. The x-axis denotes the step of training and ranges from 1 to 20000. The y-axis denotes the mean square error of the average training dataset. Thus, the convergence speed and the prediction accuracy of the training set can be obtained from the curve. It can be easily seen from the curves that the sampling density influences both the convergence speed and prediction accuracy for both networks. 

Comparing the results of two networks, three conclusions can be drawn. Firstly, when the sampling density is high enough, a higher prediction accuracy can be achieved by SIREN network than C-GRBF with enough number of parameters. It is mainly benefited from the convergence characteristic of SIREN as we mentioned. When the sampling density is less than a threshold number, the C-GRBF network can converge to a lower loss of cost function. The main reason is that C-GRBF network simulates the physics process of path impulse response formulation and thus containing prior knowledge. Secondly, the converge speed of our proposed C-GRBF neural networks is faster compared with the SIREN network when setting the same learning rate. The main reason is that the RBF layer is used to substitute several deep layers and the training algorithm for the network is back propagation. So, a shallower network with fewer parameters converge faster. The gaps of the final convergence loss between different sampling density in C-GRBF is fewer than those in the SIREN network. We guess it is the special network structure that simulates the formulation structure of CIR that makes the network less rely on the sampling density of training datasets.

\begin{figure}
	\centering
	\includegraphics[width=0.45\textwidth]{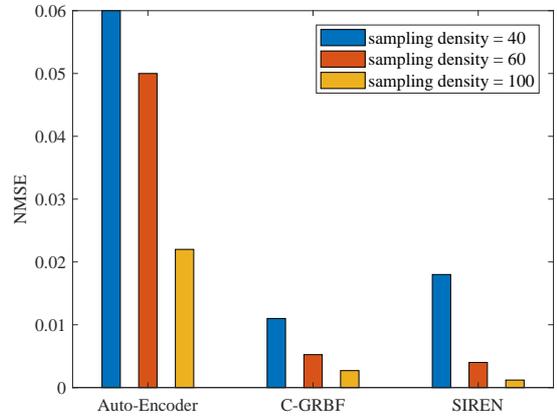}
	\caption{NMSE of testing datasets of three proposed methods with different sampling density.}
	\label{nmse}
\end{figure}

\begin{figure}
	\centering
	\includegraphics[width=0.45\textwidth]{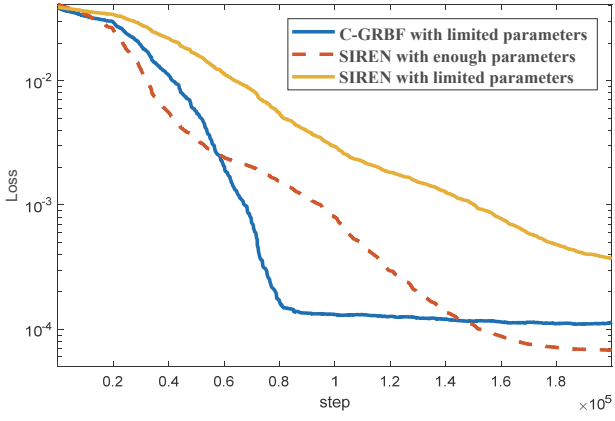}
	\caption{Convergence curve of C-GRBF and SIREN with limited parameters.}
	\label{limited_parameters}
\end{figure}

In Fig. \ref{nmse}, The performance of prediction accuracy on the testing datasets after all the learning machines are trained with enough epochs is shown. The three proposed method are compared under three sampling densities. In general, all three methods proposed by us work effectively in solving the CIR prediction problem and performs better when more training data can be accessed. Without periodic activation function, the method of Auto-Encoder performs worse compared with the other two method in whatever sampling density. In the meantime, the prediction accuracy of Auto-Encoder method decreases significantly with the decrease of the sampling density. And when the sampling density is $60$ and $100$, there exist orders of magnitude difference between the NMSE of Auto-Encoder based method and the two other networks. At the setting of sampling density=$100$, NMSE is $0.022$, $0.0027$ and $0.0012$ for AE based method, C-GRBF and SIREN respectively. At the setting of sampling density=$60$, NMSE is $0.05$, $0.0052$ and $0.0041$ respectively. At the setting of sampling density=$40$, NMSE is $0.06$, $0.011$ and $0.018$ respectively. Results of comparison between SIREN network and C-GRBF network is similar in the training dataset. SIREN network achieves a higher prediction accuracy when relatively enough training samples can be obtained. However, the C-GRBF network outperforms the SIREN network when the sampling density is lower than a certain threshold. 

\begin{figure}
	\centering
    \includegraphics[width=0.45\textwidth]{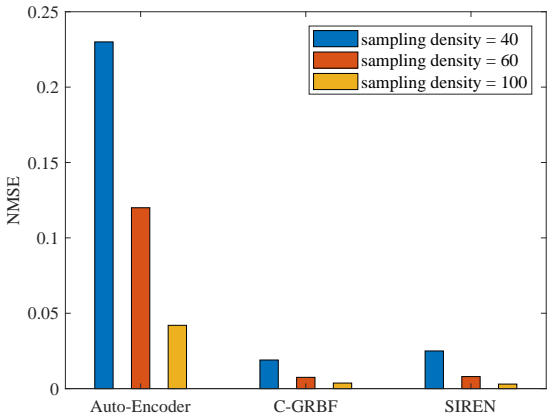}
    \caption{Prediction NMSE of testing datasets of three proposed learning structures trained through datasets with position measurement error.}
   \label{withnoise}
\end{figure}

\begin{figure}
	\centering
	\includegraphics[width=0.425\textwidth]{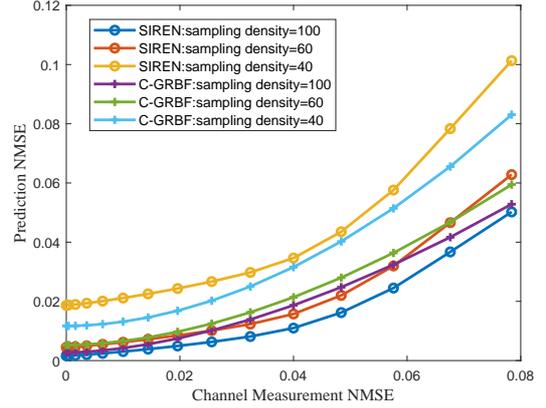}
	\caption{Prediction NMSE under different channel measurement error with Gaussian distribution in training datasets of proposed SIREN and C-GRBF network.}
	\label{robustness}
\end{figure}

\begin{figure}
	\centering
	\includegraphics[width=0.425\textwidth]{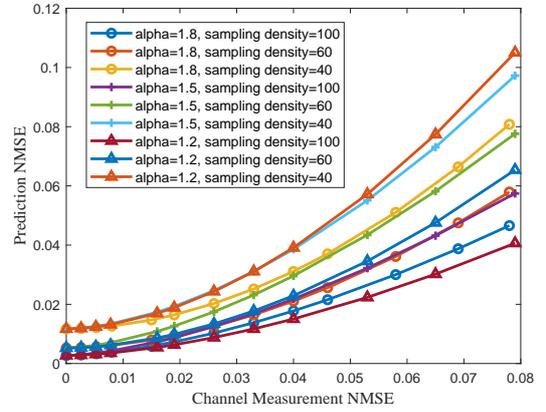}
	\caption{Prediction NMSE under different channel measurement error with different settings of symmetric alpha-stable distribution in training datasets of proposed C-GRBF network}
	\label{alphastable}
\end{figure}

\begin{figure}
	\centering
	\includegraphics[width=0.425\textwidth]{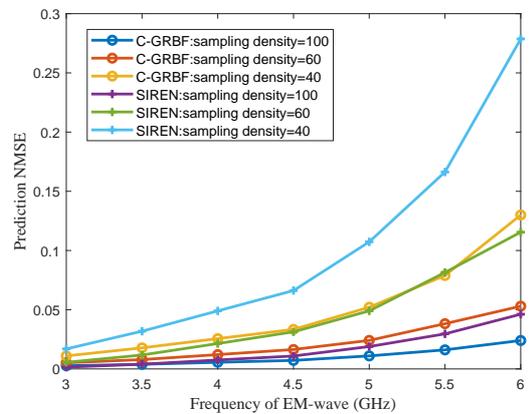}
	\caption{Prediction NMSE of proposed SIREN and C-GRBF network under different frequency band of EM-wave.}
	\label{frequency}
\end{figure}

In Fig. \ref{limited_parameters}, the results of the convergence curve of the C-GRBF and SIREN network with highly limited parameters are shown and the performance of SIREN network with enough parameters is set as the benchmark. It can be seen that with much fewer parameters, the C-GRBF network we proposed can still achieve a low convergence loss that is close to the benchmark curve while SIREN network as a universal function approximator performance with limited parameters suffers serious performance loss with fewer parameters. The results prove that the special network structure of C-GRBF network proposed by us achieves a good trade-off between the prediction accuracy and the computation and storage cost.

The robustness of three proposed methods are shown in Fig. \ref{withnoise}, Fig. \ref{robustness} and Fig. \ref{alphastable}. As mentioned in the last section, all the network parameters are kept same with the setting of Fig. \ref{nmse}. Thus, the numerical result can be directly compared. In general, all three methods proposed by us can still predict CIR with high accuracy when trained with imperfect datasets with measured position error shown by Fig. \ref{withnoise}. At the setting of sampling density=$100$, NMSE is $0.042$, $0.0037$ and $0.0031$ for AE based method, C-GRBF and SIREN respectively. At the setting of sampling density=$60$, NMSE is $0.12$, $0.0075$ and $0.0081$ respectively. At the setting of sampling density=$40$, NMSE is $0.23$, $0.019$ and $0.025$ respectively. It can be found that the robustness is directly related to the training datasets scale. When more training data is available, all the three proposed methods have stronger ability to resist the perturbation of network input. Fig. \ref{robustness} shows the comparison results of prediction accuracy with the CSI measurement accuracy in training datasets. It can be found that our proposed network still performs well when there exists relatively high measurement error in the training datasets. C-GRBF network have better robustness when the sampling density is lower, and we guess the main reason behind it is the same with what we analysed for Fig. \ref{nmse}. Fig. \ref{alphastable} shows the prediction accuracy of C-GRBF network when the channel noise is symmetric alpha-stable distributed. The main characteristic of this kind of noise distribution is that there are some extremely high interference. The results show that our proposed network have a strong capability to deal with this kind of noise distribution and the performance is quite similar with the case when the channel noise is Gaussian distributed.

In Fig. \ref{frequency}, we evaluate the prediction accuracy under different frequency band of EM-wave. All the settings are kept the same except the frequency band of EM-wave. It can be seen that due to shorter wave length, the prediction accuracy will have different degrees of decrease. However, our proposed C-GRBFnet can still retain relatively high prediction accuracy due to the special physics-inspired network design. Thus, it is promising to apply the C-GRBFnet to higher frequency band scenery with high accuracy by adding more training samples appropriately.

\begin{figure}
	\centering
	\includegraphics[width=0.425\textwidth]{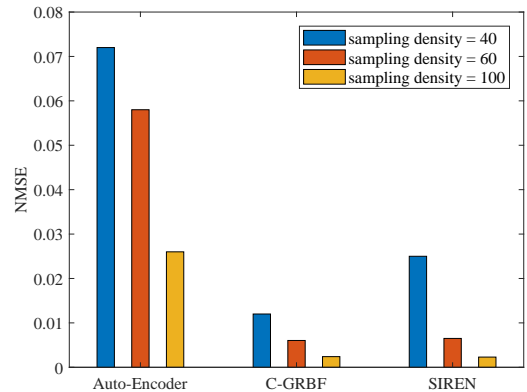}
	\caption{NMSE of testing datasets of three proposed methods with different sampling density for evaluating the performance of the proposed network when there are a lot of NLoS-only samples include.}
	\label{nmse_new_environment}
\end{figure}

In Fig. \ref{nmse_new_environment}, The performance of prediction accuracy for the scenery where a lot of NLoS-only samples are included in the dataset. The three proposed method are also compared under three sampling densities. The results are basically the same with what we shown in Fig. \ref{nmse}. The main difference is that the NMSE of our proposed C-GRBF network is lower than the SIREN network when the sampling density is $100$. And we guess the main reason is that due to the scatter environment is more complex which lead to the decrease in prediction accuracy by the fully connected neural network. By contrast, the generalization ability of C-GRBFnet is much higher and is promising to deal with more complex environment.

\begin{table}[t]
	\caption{Parameters and Values for C-GRBF and SIREN network}
	\begin{center}
		\begin{tabular}{ p{4cm}   p{4cm}}
			\toprule
			\textbf{Parameters} & \textbf{Value} \\
			\toprule
			Input dimension & 3 \\
			
			Output dimension & 64$ \times $182 \\
			
			Number of neurons in the hidden layer & C-GRBF: 256-512-1050 for the deep part and 350 kernel function in RBF layer. \\
			
			& SIREN: 512-768-1024-768-512 \\			
			
			Performance metric & Mean square error (MSE) \\
			
			Optimizer & Adam \\
			
			Training steps & $2 \times {10^5}$ \\
			
			Learning rate & $1 \times {10^{ - 3}}$ \\
			
			Batch size & 20 \\
			
			\toprule
		\end{tabular}		
	\end{center}
	\label{mimo_network_comparison}
\end{table}

\begin{table}[t]
	\caption{Parameters and Values for Channel Mapping networks}
	\begin{center}
		\begin{tabular}{ p{4cm}   p{4cm}}
			\toprule
			\textbf{Parameters} & \textbf{Value} \\
			\toprule
			Network structure & Fully connected \\
			
			Input dimension & 182$ \times $2 or 182$ \times $4 or 182$ \times $8 \\
			
			Output dimension & 64$ \times $182 \\
			
			Activation function & tanh \\
			
			Number of neurons in the hidden layer & 256-512-512-512 \\		
			
			Performance metric & Mean square error (MSE) \\
			
			Optimizer & Adam \\
			
			Training steps & $2 \times {10^5}$ \\
			
			Learning rate & $1 \times {10^{ - 3}}$ \\
			
			Batch size & 20 \\
			
			\toprule
		\end{tabular}		
	\end{center}
	\label{channel_mapping}
\end{table}

\begin{table}
	\caption{Parameters and Values for position merging LSTM networks}
	\begin{center}
		\begin{tabular}{ p{4cm}   p{4cm}}
			\toprule
			\textbf{Parameters} & \textbf{Value} \\
			\toprule
			Network structure & LSTM + FNN  \\
			
			Output dimension & 64$ \times $182 \\
			
			Activation function & tanh \\
			
		    LSTM steps & 5 \\	
			
			Neurons in LSTM hidden layer & 256 \\		
			
			Number of neurons in FNN hidden layer & 256-512-256 \\		
			
			Performance metric & Mean square error (MSE) \\
			
			Optimizer & Adam \\
			
			Training steps & $2 \times {10^5}$ \\
			
			Learning rate & $1 \times {10^{ - 3}}$ \\
			
			Batch size & 20 \\
			
			\toprule
		\end{tabular}		
	\end{center}
	\label{LSTM}
\end{table}

\section{Experimental Results For MIMO Extension} \label{chap:mimo}
\begin{figure}
	\centering
	\includegraphics[width=0.5\textwidth]{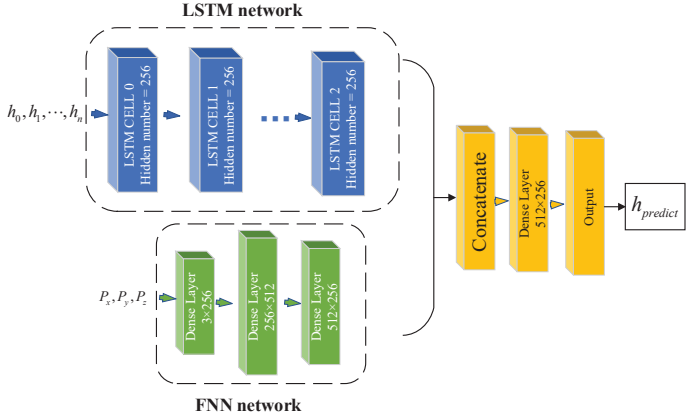}
	\caption{Network structure of merging LSTM with position information.}
	\label{mergedata}
\end{figure}

\begin{figure}
	\centering
	\includegraphics[width=0.45\textwidth]{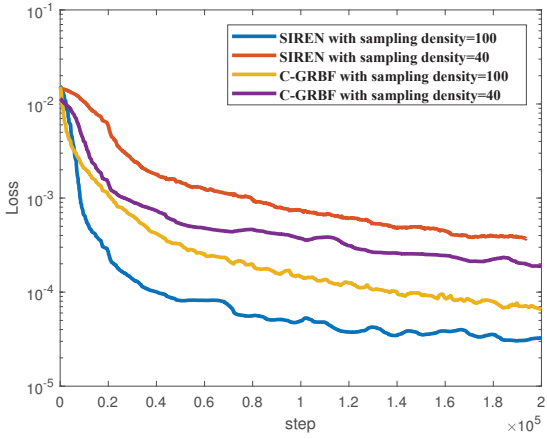}
	\caption{Convergence curve of C-GRBF and SIREN for multi-antenna system.}
	\label{multiATSIrenRBFcompare}
\end{figure}

\begin{figure}
	\centering
	\includegraphics[width=0.45\textwidth]{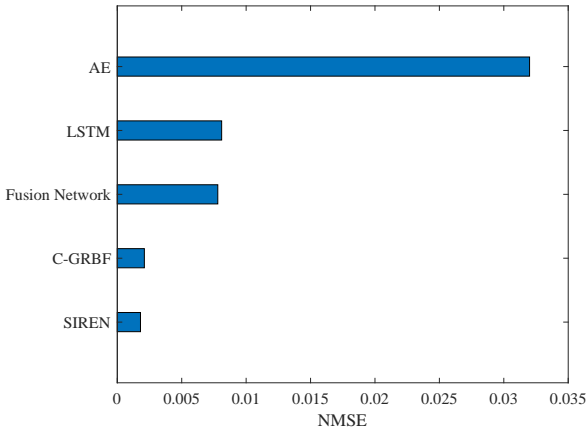}
	\caption{NMSE comparison of approaches proposed with existing fusion network.}
	\label{lstm_compare}
\end{figure}

\begin{figure}
	\centering
	\includegraphics[width=0.45\textwidth]{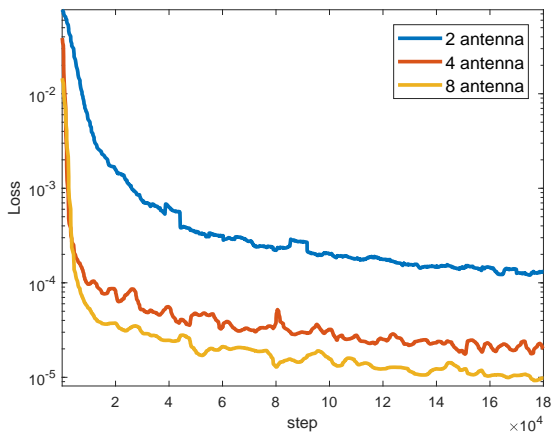}
	\caption{Convergence curve of channel mapping by 2, 4 and 8 antenna pairs.}
	\label{multiantenna_compare}
\end{figure}

\begin{figure}[t]
	\centering
	\includegraphics[width=0.45\textwidth]{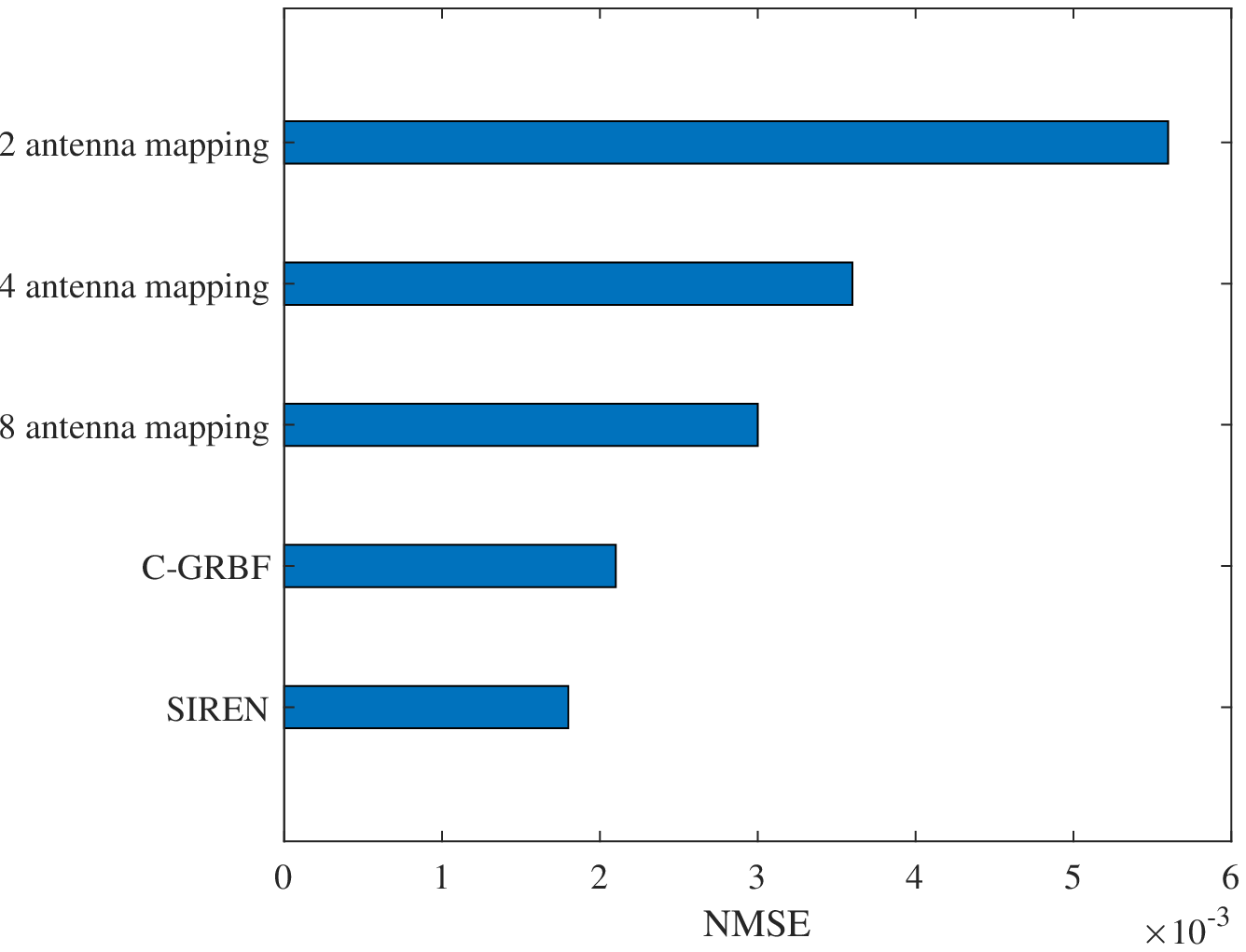}
	\caption{NMSE comparison of two approaches for extending proposed methods to multi-antenna system.}
	\label{multiATnmse}
\end{figure}

\subsection{Comparison with Existing Works}
Different from the approaches that only position information is used and wireless channel is predicted in a generative way proposed by us, most existing works that achieves relatively good performance view this problem as a sequence prediction problem. Thus, the channel sequence measured in a past period time sequence for mobile channel predicting problem or channel sequence measured in surrounding positions for static channel prediction problem are needed. In early works, Auto-regressive model is used for solving such problem. Later, LSTM neural network is found fitting these problems better. In order to directly correspond to the scenario, we only compare our approaches to those static channel predicting works with channel in several surrounding points as known factors. In \cite{9277535}, position information is also introduced to the learning machine and merged with LSTM network and the network structure is shown in Fig. \ref{mergedata}. The network is composed of three parts. The LSTM part is designed to extract feature of the sequential channel vector and the position information is added through a fully connected neural network. Then, the outputs of the hidden layer of LSTM network and fully connected network are concatenated to make up a vector and the predicted channel is outputted by a fully connected network. It is worth noting that although both approaches proposed by us and work in \cite{9277535} need to use some sampling channel in a fixed area, the data obtain process is quite different due to the difference of prediction mechanism and network structure. As a sequence prediction problem, this work needs to sample points in a straight line with equal distance for predicting each single channel in a certain position while our approach only needs to use historical data with random sampling position as mentioned above. Thus, our approaches are far more flexible. Experiments are conducted in the following part to compare the prediction accuracy with a similar scale of neural network and training data scale.

\subsection{Experiment Setting and Simulation Results}
\subsubsection{Experiment Setting}
The experiment is conducted in the same scene illustrated in section \ref{chap:datasets} and the single antenna is replaced with a $8 \times 8$ planar antenna array. The planar antenna array is set to be parallel with x-y plane. The position coordinates of the first element of the array is set to be ($45$m,$18$m,$37$m) and the interval between elements is half wave length. The channel mapping by 2, 4 and 8 antennas are evaluated in our work. At last, the NMSE of the prediction results crossing the testing datasets is compared between those methods. The detailed parameters and values are summarized in Table \ref{mimo_network_comparison} and Table \ref{channel_mapping}.

We build the training and testing datasets for the comparing methods in the same scene but with different sampling mode. Corresponding to the sampling method mentioned in \cite{9277535}, we sample channel CIR sequence along the y-axis. The sequence length for a training data is set to be 5 and the distance between two adjacent sampling points is set to be 0.1m for the reason that 0.1m is the average distance in our work when the sampling density is set to be 100. For the fairness of comparison, the number of the whole dataset is set to be the same with our works. Thus, the whole number of sampling points is 5 times of our works. The fusion network structure proposed in \cite{9277535} is shown in Fig. \ref{mergedata} and detailed parameters and values are summarized in Table \ref{LSTM}. In order to explore the effectiveness of the fusion network structure, we also try to directly omit the fully connected part for adding position information and only the LSTM part is retained.

\subsubsection{Simulation Results}
As shown in Fig. \ref{multiATSIrenRBFcompare}, experiments with two sampling density show that the performance of SIREN and C-GRBF methods can still achieve low MSE in multi-antenna systems. And the characteristics of the four curves are similar to what we analyzed in the single antenna scene. Basically, when the sampling density is lower, the proposed C-GRBF network can achieve higher prediction accuracy than SIREN.

Through analyzing the results shown in Fig. \ref{lstm_compare}, we find that SIREN and C-GRBF network proposed by us can achieve higher prediction accuracy than the comparing work without taking measured channel as input of the network. The results of the fusion network and the simple LSTM network show that the added position information does not play an effect for the prediction task which means this network structure does not have the ability of scattering environment learning. As we analyzed above, LSTM network only extracts the channel correlation of adjacent sampling points while the network proposed by us learns the implicit representation of scattering environment and mechanism of electromagnetic propagation, which makes our network more flexible and practical.

Fig. \ref{multiantenna_compare} shows the capability of channel mapping. It is shown that it is reliable to use the idea of channel mapping and higher accuracy can be achieved by using more antenna channel as the input at the expense of higher computation cost. Here, it should be noted that the input of the channel mapping neural network is the recorded measured channel CIR. In practical usage, the prediction accuracy will get some loss by using the predicted CIR of several single antennas as input because of the accumulation of prediction error.

The performance of predicting accuracy of testing datasets is shown in Fig. \ref{multiATnmse}. The input of channel mapping networks is the predicted CIR vector of single antenna of SIREN network. Both SIREN and C-GRBF can achieve good prediction accuracy and the final NMSE value is close to what in the single antenna scenario. Caused by error accumulation, NMSE of the channel mapping approach is higher but still low enough for practically application.

\section{CONCLUSION} \label{chap:conclusion}
In this paper, we solved an undiscussed problem of efficiently and accurately predicting the static channel impulse response with only the user's position information and a set of channel instances obtained within a certain wireless communication environment. Innovatively, the problem was viewed as a generative problem and a data-driven way is proposed for solving it. We illustrated why universal neural network with increasing activation function cannot handle such a problem by analyzing the characteristic of mathematical expression of CIR and the mechanism of function approximation by universal neural network. Then, a novel physics-inspired generative learning structure to implicitly represent and predict the static channel characteristics was proposed by us. Furthermore, two possible solutions to extend our methods to MIMO system are proposed. Experiments had shown the effectiveness of our proposed learning structures. The values predicted through our method were very close to the original data generated by Ray-Tracing model.

One inspiring achievement of our work is that we did a meaningful attempt to show that the wireless communication scenario can be learned by physics-inspired learning structures and algorithms. Thus, precisely designing and optimizing the wireless communication system in an intelligent way is possible in the future. Our work is a good example to illustrate how to design specific learning structure to adapt to a special problem in wireless communication. Moreover, our novel learning structure works significantly better than those universal learning structure and have the potential to guide the learning system design of sensing wireless communication scene and intelligent wireless communication in the future.

\bibliographystyle{IEEEtran}
\bibliography{bibfile}

\end{document}